\documentclass[twocolumn]{revtex4}
\usepackage[dvips]{graphicx}

\begin{document}

\bibliographystyle{unsrt}

\title{Turning Point Prediction of Oscillating Time Series using Local Dynamic Regression Models}

\author{Dimitris Kugiumtzis}
\email{dkugiu@gen.auth.gr}
\author{Ioannis Vlachos}
\email{ivlaxos@gen.auth.gr}

\affiliation{Department of Mathematical, Physical and Computational Sciences \\
Faculty of Technology, Aristotle University of Thessaloniki, Greece
}

\begin{abstract}
In the prediction of oscillating time series, the interest is in the turning points of successive oscillations rather than the samples themselves. For this purpose a scheme has been proposed; the state space reconstruction is limited to the turning points and the local (nearest neighbor) model is modified in order to predict the turning point magnitudes and times. This approach is extended here using a local dynamic regression model on both turning point magnitudes and times. Simulations on oscillating nonlinear systems show that the proposed approach gives better predictions of turning points than the standard local model applied to all the samples of the oscillating time series.
\end{abstract}

\maketitle

\section{Introduction}
Oscillating time series are common in applications and are
characterized by series of patterns of an upward trend followed
by a downward trend. When oscillating time series do not exhibit apparent periodicity, such as those generated by chaotic systems, the problem of prediction lies basically on the time and magnitude of the peaks and troughs, as the results of three time series competitions showed \cite{Weigend94,Suykens98,ESTSP07}. Interestingly, the
winning prediction schemes in the two first competitions were
based on a local prediction model (with rather involved
modifications of the standard nearest neighbor prediction
approach).
Local models stem from dynamical systems and chaos theory, are computationally efficient, and perform as well as more
complicated black-box models, such as neural networks, in the
prediction of irregular time series \cite{Kantz97}. For multi-step ahead prediction typically a higher embedding dimension $M$ is required. For a fixed time delay $\tau$, the reconstructed points span a time window of length $\tau_w=(M-1)\tau$. This should be large enough to account for the mean orbital period of the underlying trajectory, i.e. $\tau_w$ should cover the period of an oscillation or a pattern of oscillations \cite{Kugiumtzis96}.


Turning point prediction is of great practical interest in many applications, such as finance \cite{Bao08}. A recently developed approach attempts to model oscillating time series from low-dimensional systems with the so-called peak-to-peak dynamics \cite{Piccardi08}. This approach relies on simple one or two dimensional maps for the peaks. In \cite{Kugiumtzis08b}, it was shown that the prediction of
turning points with local models is improved using state space
reconstruction on the time series of turning points at a lower embedding dimension $m$. Here, we extend the state space reconstruction to include also the time series of the times of the turning points. This is the setting of local dynamic regression, where a local model on two time series (for magnitudes and times of turning points) is build in order to predict the magnitudes and times of turning points.

\section{State Space Reconstruction of Turning Points}

\begin{figure}[h!]
\hspace{7mm} \includegraphics[height=35mm]{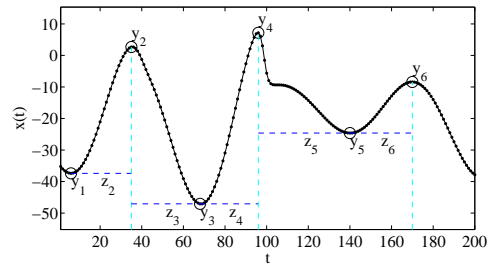}
\caption{Time series of turning point magnitudes and trend durations derived from an oscillating time series.}
 \label{fig:oscext}
\end{figure}

Suppose an oscillating time series of length $N$,
$\{x(t)\}_{t=1}^N$, is observed at a sampling time $\tau_s$. A sample $y_i=x(t_i)$ is a turning point of $\{x(t)\}_{t=1}^N$ at time step $t_i$ if it is the minimum or maximum of all samples in $[t_i-p,t_i+p]$, for a scale parameter $p$. Scanning all samples of $\{x(t)\}_{t=1}^N$, the time series $\{y_i\}_{i=1}^n$ and $\{t_i\}_{i=1}^n$ of magnitudes and times of the alternating turning points, respectively, are derived.
Instead of the time of the turning points we derive the duration of the upward and downward trends from the first differences $z_i=t_i-t_{i-1}$, giving the time series $\{z_i\}_{i=2}^n$.
Thus two successive samples of $\{y_i\}_{i=2}^n$ together with the synchronous samples of $\{z_i\}_{i=2}^n$ regard an oscillation of $\{x(t)\}_{t=1}^N$, as shown in Fig.~\ref{fig:oscext}.
The bivariate time series $\{y_i,z_i\}_{i=2}^n$ compresses the information in $\{x(t)\}_{t=1}^N$ with some loss of information depending on the pattern of the samples between the turning points. At the limit where the upward and downward trends are linear there is no loss of information as any sample $x(t_i-k)$ between two turning points $x(t_{i-1})$ and $x(t_i)$, where $k \in \{0,1,\ldots,t_i-t_{i-1}\}$, can be expressed in terms of the magnitude and time of the two turning points as
\[
x(t_i-k) = x(t_i)-k\frac{x(t_i)-x(t_{i-1})}{t_i-t_{i-1}}=
 y_i-k\frac{y_i-y_{i-1}}{t_i-t_{i-1}}.
\]

The state space reconstruction of $\{y_i\}_{i=2}^n$ can be considered as a specific state space reconstruction of $\{x(t)\}_{t=1}^N$ at time points $\{t_i\}_{i=1}^n$ for delays depending at each $t_i$. For an embedding dimension $m$, this reads
\begin{equation}
 \begin{array}{rcl}
 \mathbf{y}_i & = & [y_i,y_{i-1},\ldots,y_{i-m+1}]^{\prime} \\
 & = & [x(t_i),x(t_{i-1}),\ldots,x(t_{i-m+1})]^{\prime},
 \end{array}
 \label{eq:embedextpoi}
\end{equation}
for $i=m,\ldots,n$ \cite{Kugiumtzis08b}. The advantage of this reconstruction is that it reduces the embedding dimension $M$ of the standard state space reconstruction of the type $\mathbf{x}(t)=[x(t),x(t-\tau_1),\ldots,x(t-\tau_{M-1})]^{\prime}$ to $m$. Usually, in prediction tasks the delays $\tau_j$ are small (and commonly a fixed delay $\tau$ is used) suggesting a rather large $M$ in order the time window $\tau_w$ to cover the mean oscillation period.

We extend the state space reconstruction in (\ref{eq:embedextpoi}) to account for the duration of trends. The analysis on the bivariate time series $\{y_i,z_i\}_{i=2}^n$ requires that both time series are standardized (subtracting the mean and dividing with the standard deviation for each time series). The state space reconstruction on the standardized $\{y_i,z_i\}_{i=2}^n$ reads
\begin{equation}
 \mathbf{w}_i = [y_i,y_{i-1},\ldots,y_{i-m_y+1},z_i,z_{i-1},\ldots,z_{i-m_z+1}]^{\prime}. \label{eq:embedmagtim}
\end{equation}
We allow for different embedding dimensions $m_y$ and $m_z$ for the magnitudes of turning points and duration of trends, respectively.

\section{Dynamic Regression Prediction of Turning Points}

The prediction of $y_{i+T}$ and $z_{i+T}$ for a lead time $T$ can be posed independently and this constitutes a problem of dynamic regression (also termed distributed lag modeling) \cite{Pankratz91}. In this setting we apply local average models (LAM) and local linear models (LLM) \cite{Kantz97}. The prediction of $y_{i+T}$ with LAM is given by the average of the $T$ step ahead mappings of the $K$ nearest neighboring points to $\mathbf{w}_i$
$\{\mathbf{w}_{i(1)},\ldots,\mathbf{w}_{i(K)}\}$
\begin{equation}
 \hat{y}_{i+T} = \bar{y}_{i(K)+T} = \frac{1}{K}\sum_{j=1}^K y_{i(j)+T}.
 \label{eq:lammag}
\end{equation}
Assuming a linear autoregressive model restricted to the neighboring points to $\mathbf{w}_i$, the LLM prediction of $y_{i+T}$ is
\begin{equation}
 \hat{y}_{i+T} = \bar{y}_{i(K)+T} + \mathbf{a}^{\prime} (\mathbf{w}_i - \bar{\mathbf{w}}_{i(K)}),
 \label{eq:llmmag}
\end{equation}
where $\bar{\mathbf{w}}_{i(K)}$ is the center point of the $K$ neighboring points to $\mathbf{w}_i$ and $\mathbf{a}$ is estimated from the minimization of the error function
\begin{equation}
\sum_{j=1}^K \left(y_{i(j)+T}-\bar{y}_{i(K)+T}-\mathbf{a}^{\prime} (\mathbf{w}_{i(j)} - \bar{\mathbf{w}}_{i(K)})\right)^2.
 \label{eq:errfun}
\end{equation}
We consider also regularization of the ordinary least square solution of (\ref{eq:errfun}) making use of principal component regression (PCR) and projection on the $q$ first components \cite{Kugiumtzis98}. Note that $z_{i+T}$ is predicted in the same way, but in line with dynamic regression setting the suitable embedding dimensions $m_y$ and $m_z$ may be different for the models (LAM or LLM) for $y_{i+T}$ and $z_{i+T}$.
This approach differs from the approach in \cite{Kugiumtzis08b} in that the neighboring points are formed not only based on the turning point magnitudes but also on the duration of trends.
Both LAM and LLM models are simple extensions of the respective local models used for univariate time series. Note that the direct scheme is used here, but the iterative prediction scheme can be applied in a similar way (in \cite{Kugiumtzis08b} it was found that the iterative scheme of LAM on $\{x(t)\}_{t=1}^N$ gave worse results). In the following, we compare the prediction of turning points (magnitude and time) using LAM or LLM models estimated on all the samples of the oscillating time series $\{x(t)\}_{t=1}^N$ (denoted osc-LAM and osc-LLM) and on the bivariate time series of turning point magnitudes and trend durations $\{y_i,z_i\}_{i=2}^n$ (denoted tur-LAM and tur-LLM).

\section{Turning Point Prediction on Simulated Systems}

Before presenting the results of the predictions on selected simulated systems we make some general observations regarding the implementation of the prediction schemes. For a fixed number of oscillations, $N$ is inversely proportional to $\tau_s$, so that a better time resolution of the measurements implies a larger oscillating time series $\{x(t)\}_{t=1}^N$, whereas the length of the turning point time series $\{y_i,z_i\}_{i=2}^n$ is unaffected (being $2(n-1)$).
A small $\tau_s$ is actually welcome in the analysis based on turning points because it allows for more accurate detection of the turning points and especially the trend durations. For example, for a time series with an average oscillation period of 10 samples the range of $z_i$ is limited to integers from 1 to less than 10, and this range is insufficient to define neighborhoods (in the projected reconstructed state space of dimension $m_z$). Thus a smaller $\tau_s$ would render the information of $\{z_i\}_{i=2}^n$ more useful in the setting of dynamic regression.

The parameter $p$ that defines the local window for the detection of turning points depends on $\tau_s$ and should not be too large, so that turning points of short lasted oscillations can be detected, and not too small, so that glitches of noisy oscillations are not assigned to turning points. For the latter case, a small $p$ can still be used if the time series is filtered, and then the turning points are detected on the smoothed time series to give the turning point times, whereas the turning point magnitudes are taken from the original time series. In the simulations we use $p=3$ and filter only noisy data with an order depending on the system and noise amplitude.

\begin{figure}[h!]
\hspace{7mm} \includegraphics[height=35mm]{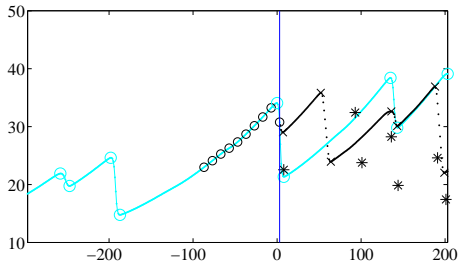}
\caption{Turning point prediction: the real time series segment (grey lines and circles), the sample and turning point predictions with osc-LAM (dark dots and crosses), and the turning point prediction with tur-LAM (dark asterisks). The current time of the turning point is set to 0 and the delays of the standard embedding on the samples are shown with open circles.}
 \label{fig:preoscext}
\end{figure}

When predicting turning points with osc-LAM or osc-LLM on $\{x(t)\}_{t=1}^N$, the current point is not a turning point $x(t_i)$ but the sample $x_{t_i+p}$ at the time the turning point $x(t_i)$ can first be detected. Thus using a large $p$ or $\tau_s$ favors the prediction on $\{x(t)\}_{t=1}^N$ because then the current point is well in the next trend of the oscillation. The prediction schemes are illustrated in Fig.~\ref{fig:preoscext}.
Note that the predicted turning points with osc-LAM are detected among the multi-step sample predictions in the same way as the turning points are detected in the oscillating time series.

\begin{figure}[h!]
\centering
\hbox{\includegraphics[height=33mm]{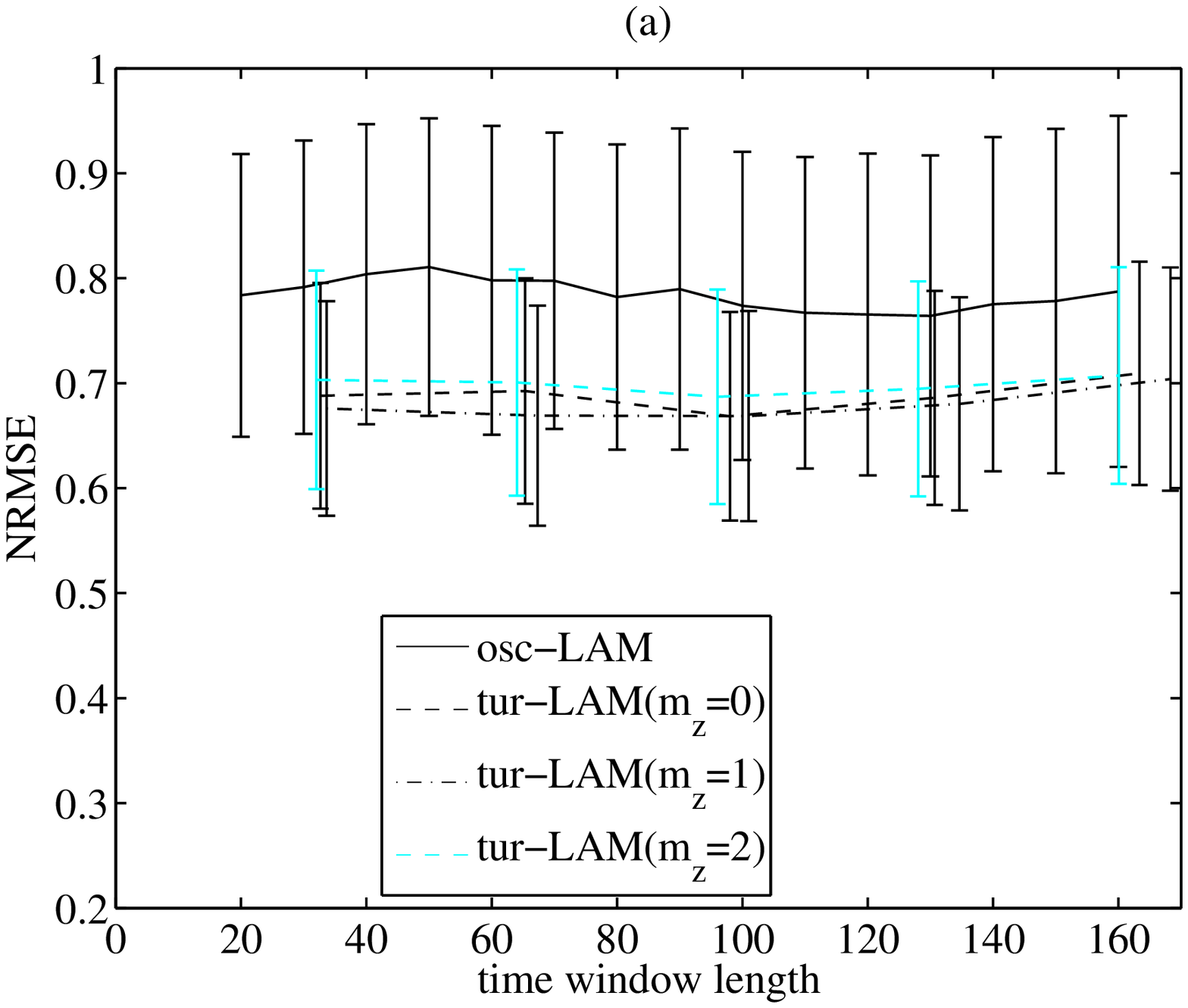}
\hspace{7mm} \includegraphics[height=33mm]{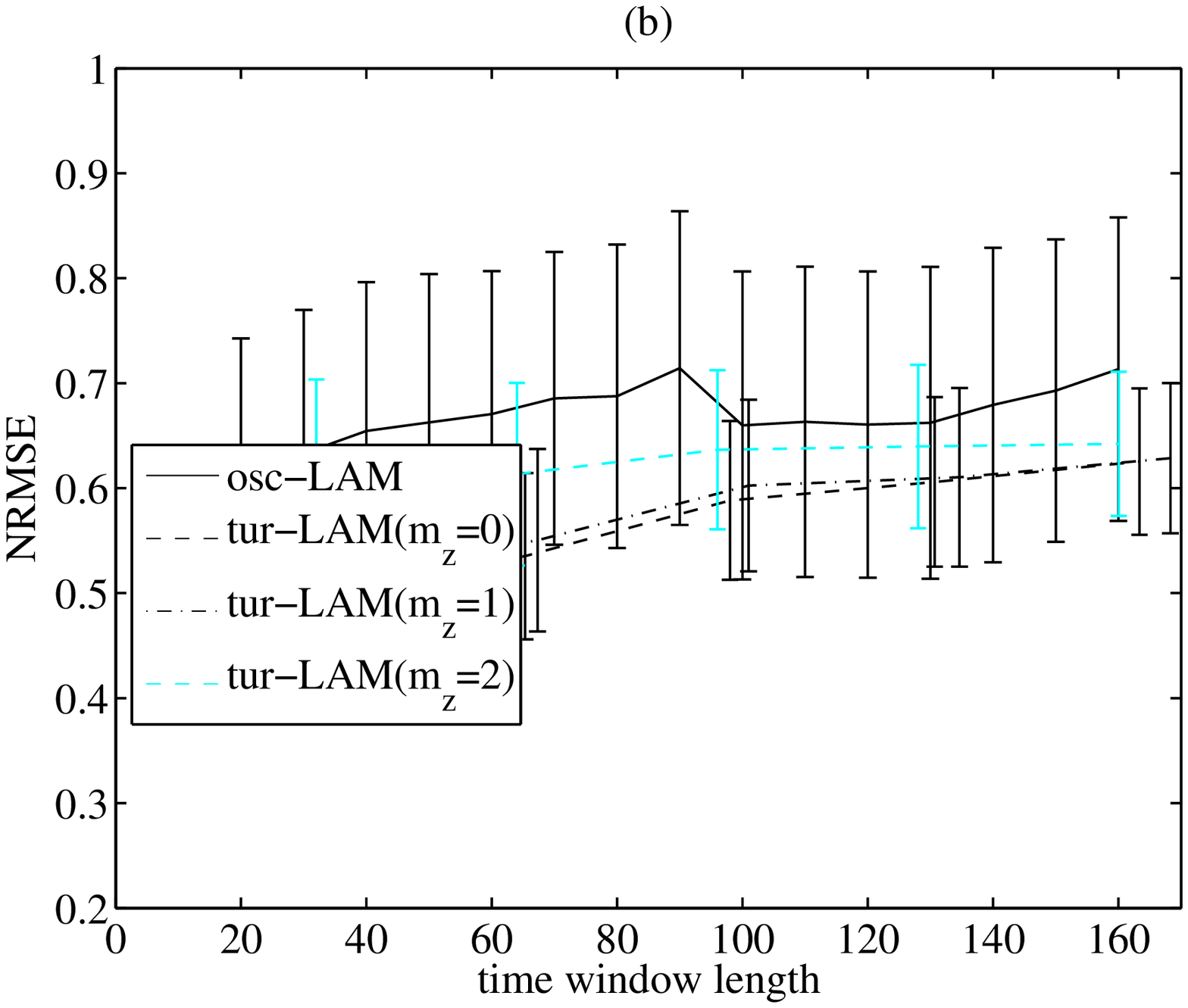}}
\hbox{\includegraphics[height=33mm]{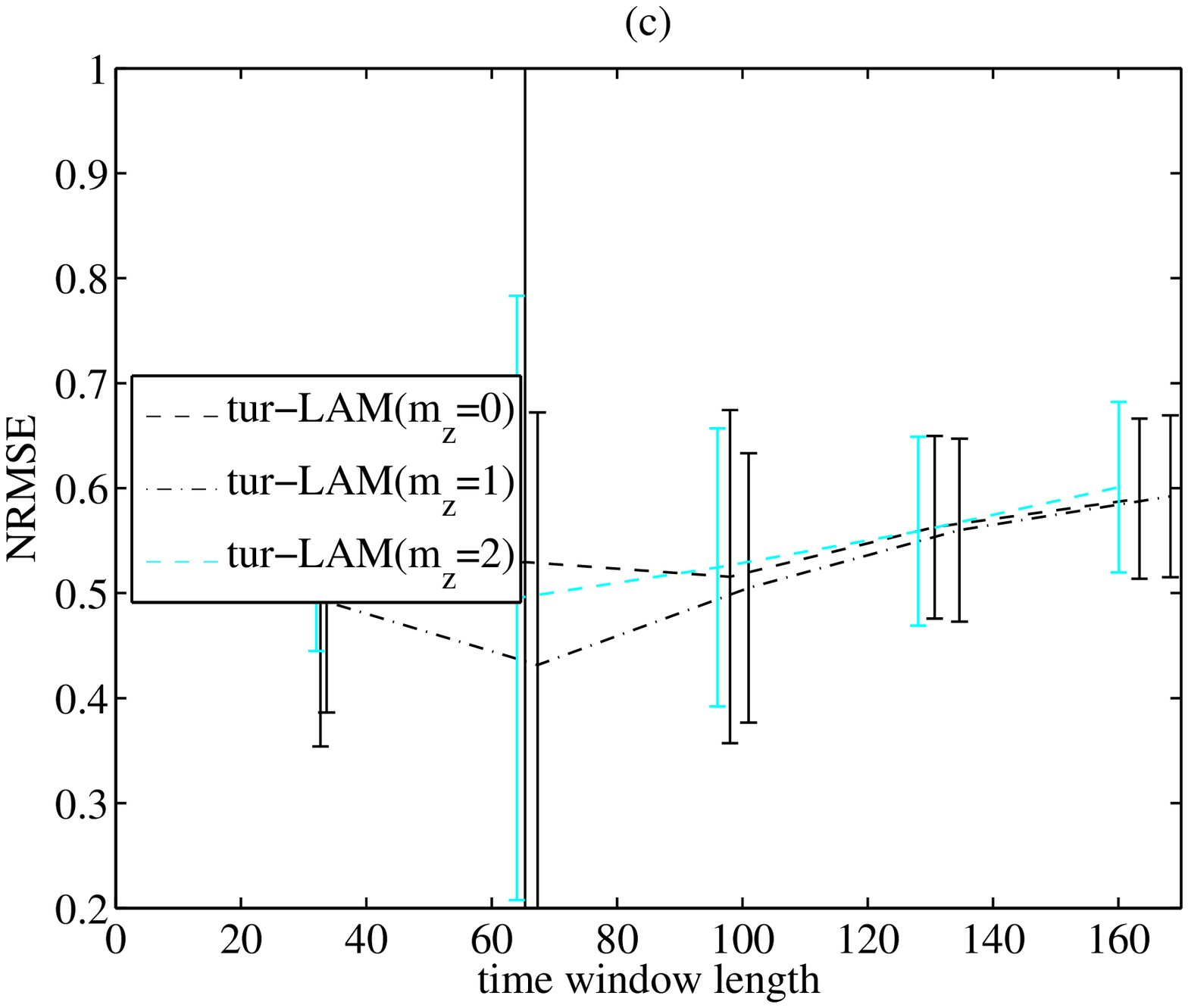}
\hspace{7mm} \includegraphics[height=33mm]{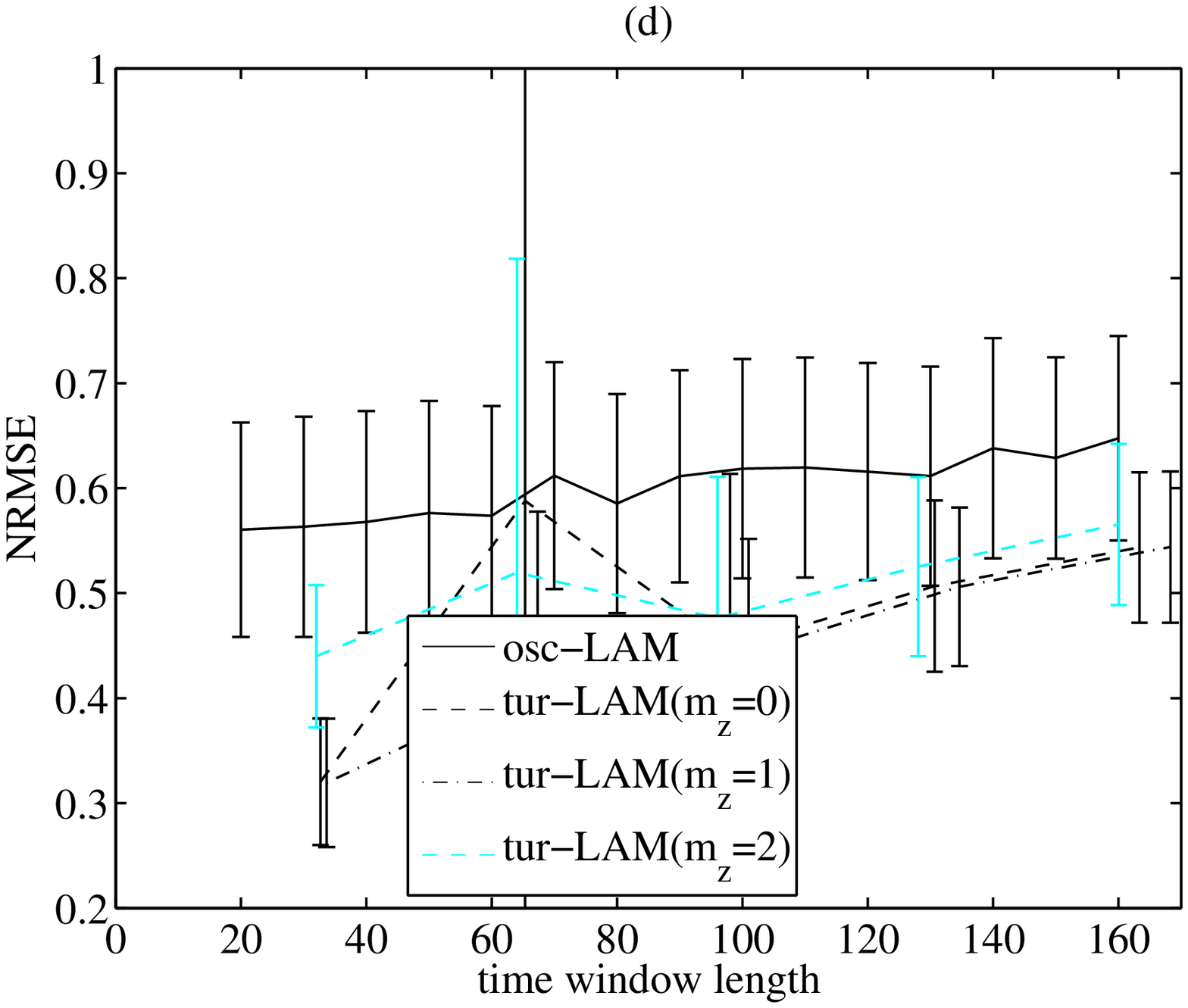}}
\caption{(a) The average NRMSE (with error bars for the standard deviation) of the prediction of next turning point magnitude of the fourth variable of the R\"{o}ssler hyperchaos system ($\tau_s=0.1$, $N=2^{14}$) using osc-LAM and tur-LAM (for $m_z=0,1,2$ as given in the legend). The $\tau_w$ in the abscissa is defined as $\tau_w=(M-1)10$ for osc-LAM and $\tau_w=(m-1)33$ for tur-LAM, as the mean oscillation period is estimated to be 66. (b) As in (a) but for the prediction of trend duration. (c) and (d) are as (a) and (b), respectively, but using the LLM models instead with PCR regularization parameter $q=3$.}
 \label{fig:MCHyp}
\end{figure}

We applied the LAM and LLM schemes on multiple realizations of known systems, such as the first and third variable of the R\"{o}ssler system \cite{Roessler76}, the first and fourth variable of the R\"{o}ssler hyper-chaos system \cite{Roessler79} (a segment of this is shown in Fig.~\ref{fig:preoscext}), and the Mackey-Glass delay differential equation for different delays $\Delta=17,30,100$ \cite{Mackey77}. The prediction measure is the normalized root mean square error (NRMSE) of the turning point prediction at the last quarter of each time series. In Fig.~\ref{fig:MCHyp}, the average NRMSE (with the standard deviation forming the error bars) is shown for 1000 realizations of the fourth variable of the R\"{o}ssler hyper-chaos system using the osc-LAM and tur-LAM models as well as the osc-LLM and tur-LLM models.
The parameters of state space reconstruction for both $\{x(t)\}_{t=1}^N$ and $\{y_i,z_i\}_{i=2}^n$ were chosen so that $\tau_w$ covers up to three mean oscillation periods. For the latter, different combinations of $m_y$ and $m_z$ were considered and in Fig.~\ref{fig:MCHyp} the tur-LAM and tur-LLM are shown for $m_z=0,1,2$ ($m_z=0$ denotes that the model is built only on the turning point magnitudes). In this example, there is little improvement of turning point prediction using the trend durations. Using either LAM or LLM, the prediction of turning points based on $\{y_i,z_i\}_{i=2}^n$ is superior. For osc-LLM prediction of turning point magnitudes, NRMSE is larger than one (the mean prediction) and has a large variance (not shown in Fig.~\ref{fig:MCHyp}c). The linear mapping diverges for multi-step ahead predictions, and we conjecture that this is because temporally close points are selected in the set of the $K$ neighboring points. The large variance of NRMSE is observed with all LLM models for $m=3$ (equal to $q$) and this needs further investigation.


The best predictions of LAM for turning point magnitudes and trend durations for both $\{x(t)\}_{t=1}^N$ and $\{y_i,z_i\}_{i=2}^n$ are given in Table~\ref{tab:MCHyp}.
\begin{table}[h!]
  \centering
   \begin{tabular}{|c|c|cc|ccc|}
    \hline
    \multicolumn{2}{|c|}{} & \multicolumn{5}{c|}{Turning point magnitude} \\ \hline
    $T$ & $K$ & $M$ & NRMSE & $m_y$ & $m_z$ & NRMSE \\ \hline
 1 & 1 & 9 & 0.604 & 3 & 1 & 0.505 \\
 1 & 5 & 9 & 0.621 & 3 & 1 & 0.518 \\
 1 & 10 & 9 & 0.662 & 2 & 1 & 0.558 \\
\hline
 2 & 1 & 10 & 0.837 & 3 & 1 & 0.679 \\
 2 & 5 & 8 & 0.748 & 3 & 1 & 0.642 \\
 2 & 10 & 3 & 0.732 & 3 & 0 & 0.665 \\
\hline
    \multicolumn{2}{|c|}{} & \multicolumn{5}{c|}{Trend duration} \\ \hline
 1 & 1 & 10 & 0.669 & 2 & 1 & 0.368 \\
 1 & 5 & 3 & 0.549 & 2 & 0 & 0.366 \\
 1 & 10 & 3 & 0.526 & 2 & 0 & 0.414 \\
\hline
 2 & 1 & 10 & 1.016 & 3 & 1 & 0.817 \\
 2 & 5 & 10 & 0.989 & 2 & 0 & 0.772 \\
 2 & 10 & 9 & 1.018 & 2 & 0 & 0.782 \\
\hline
  \end{tabular}
  \caption{For the system in Fig.~\ref{fig:MCHyp} and for each combination of $T=1,2$ and $K=1,5,10$, the $M$ of best prediction with osc-LAM and $m_y$ and $m_z$ of best prediction with tur-LAM together with the respective NRMSE are given, where $M=3,\ldots,10$ ($\tau=10$) and $m=2,\ldots,6$.}
  \label{tab:MCHyp}
\end{table}
The best turning point predictions (magnitude and time) are derived with tur-LAM at small embedding dimensions (up to 3 for $m_y$ and 0 or 1 for $m_z$). Closer investigation showed that for some prediction tasks osc-LAM predicted better than tur-LAM, whereas in other cases it formed a turning point far from the true turning point, so overall the NRMSE was worse. This difference in osc-LAM and tur-LAM persists for different $N$ (we tested for $\log_2N=12,13$) and at a lesser extent also for the addition of observational noise (we used $5\%$ and $10\%$ of noise amplitude). Moreover, the inclusion of the last trend duration ($m_z=1$) improved the prediction of turning point magnitudes and only marginally the prediction of trend durations. The same qualitative results were obtained from the same simulations on the other systems. For the highly complex Mackey-Glass system with $\Delta=100$ (it has a fractal dimension of about 7) the best results of tur-LAM were obtained for high embedding dimensions of both turning point magnitudes and trend durations, indicating that for this system the trend duration is important to predict the next peaks or troughs.

\section{Conclusion}
We showed that the local prediction of turning points (magnitude and time) can be improved if the nearest neighbor model of average or linear mapping is build on reconstructed points from the bivariate time series of turning point magnitudes and trend duration.



\section*{Acknowledgments}
The work is part of the research project 03ED748 within the framework of the ''Reinforcement Programme of Human Research Manpower'' (PENED) and it is co-financed at 90\% jointly by European Social Fund (75\%) and the Greek Ministry of Development (25\%) and at 10\% by Rikshospitalet, Norway.

\bibliography{c:/MyFiles/Papers/LaTeX/ChaosReferences}


\end{document}